\documentclass{article}[12pt]

\newcommand{\be}{\begin{equation}} 
\newcommand{\ee}{\end{equation}}

\begin{document}

\begin{titlepage}

\rightline{EFI-01-18} \rightline{hep-th/0105219}

\begin{center}

\vskip 3cm

\Large{\bf

Inflation and Holography in String Theory}

\large

\vskip 2cm Bruno Carneiro da Cunha\footnote{bcunha@theory.uchicago.edu}

\vskip 12pt {\sl Enrico Fermi Inst. and Dept. of Physics \\ The
University of Chicago \\  5720 S. Ellis Ave., Chicago IL 60637, USA}

\normalsize

\vspace{2cm}

\end{center}

\abstract{ The encoding of an inflating patch of space-time in terms
of a dual theory is discussed. Following Bousso's interpretation of
the holographic principle, we find that those are generically
described not by states in the dual theory but by density matrices. We
try to implement this idea on simple deformations of the AdS/CFT
examples, and an argument is given as to why inflation is so elusive to
string theory.  }

\end{titlepage}

\input{psbox}

\newpage

\large

\section{Introduction}

The idea that the area of a surface in space-time bounds the number of
degrees of freedom in the space-like region inside it has been
entertained for some time now \cite{'thooft}. The quintessential
example of it is the semi-classical black hole, whose entropy scales
as its area and is in some sense the most entropic system one can
imagine \cite{susskind}.

Some realizations of this have arisen in string theory, particularly
in the $\mbox{AdS}/\mbox{CFT}$ correspondence. Although there the
theory constructed is dual to gravity, one can nonetheless estimate
that the number of degrees of freedom inside a bounded region is
proportional to the area of its boundary \cite{susskwitten}. The
theory thus constructed - the CFT - can then be sensibly described as
living ``on the boundary of space-time.''

This holographic description of space was then generalized by Bousso
\cite{bousso}, who succeded in giving a covariant  set of rules to
associate regions of space-time with the degrees of freedom living on
a surface bounding it. One of the strengths of this approach is the
transparency of the statistical mechanical interpretation of the
entropy, since the rules are independent of the ``arrow of time'' one
chooses. Of particular interest for this construction are the
horizons, null surfaces where the boundary of an observable patch of
space-time lies. As such, the holographic description of points on
either side of the horizon can be located at distant regions of space
(see below).

This work addresses the problem of horizon formation in the the
supergravity theories that arise from string theory. The motivation
seems rather clear, given the amount of information gathered recently
from the interplay between supergravity and string theory. More
prosaically, this situation seems to combine both ingredients
necessary \cite{halliwell}: i)  Kaluza-Klein compactifications can
generate exponential potentials that are on the verge of inducing
inflation, and ii) Even though the antisymmetric tensor fields of
supergravity obey the strong energy condition in 10 (or 11)
dimensions, this may not be so after compactification.

The paper is organized as follows. The second section reviews Bousso's
contruction - and terminology - of preferred screens and the context
of formation of horizons. The third and fourth sections deal with
supergravity compactifications. it finishes with a discussion about
the results and interpretation.

\section{Holographic Screens}

Loosely speaking, holography is the statement that the area of a
surface bounds the number of degrees of freedom in its interior. Of
course, this idea has a serious drawback that there is no covariant
notion of ``interior region'' when one is dealing with a $n-2$
hypersurface embedded in a $n$ dimensional space-time. This notion can
be recovered \cite{bousso} by considering collapsing null light rays
emanating from the surface, the {\em light-sheets}. This gives then a
covariant definition of the interior region described by its boundary
degrees of freedom.

By following  the geodesic generators  of the light-sheet back  in the
direction of non-negative expansion, one can {\em project} the degrees
of freedom  into a  larger, possibly infinite,  surface $B$,  the {\em
screen}. Generically, the  expansion will change sign at  $B$, but, if
the  expansion vanishes everywhere  in $B$  it will  be called  a {\em
preferred screen}.  Preferred screens are  conjectured to be  those in
which the holographic bound is saturated.

One  can then  foliate the  whole space-time  into  null-surfaces, and
project their degrees  of freedom onto screens. The  family of screens
contains a finite number of degrees of  freedom per Planck area. By an
abuse of language we will also call these hypersurfaces preferred
screens. Examples of preferred screens include the cosmological
horizon of an inflationary universe and the conformal boundary of an
anti-de Sitter (AdS) space-time.

Consider for instance the maximally extended Schwarzchild-AdS space
(Fig. 1): there are two preferred screens, at the conformal boundaries
of space-time. By null projection, one  can encode all the information
inside the black-hole into either boundary.  For instance, a state
localized inside the horizon of the black hole, at for instance a
point $P$ can be described by the set of degrees of freedom in the
boundary as a state in the Hilbert space ${\cal H}_1$ living at the
boundary preferred screen. This view is compatible with the AdS/CFT
duality, which  states that  the (conformal) field theory describing
perturbations  of  AdS  space  lives at  the conformal boundary. Note
however that the local measurements on $P$ do depend on information
coming from both regions near the spatial infinities.

\begin{figure}[ht]
\begin{center}
\mbox{\psannotate{\psboxto(8cm;0cm){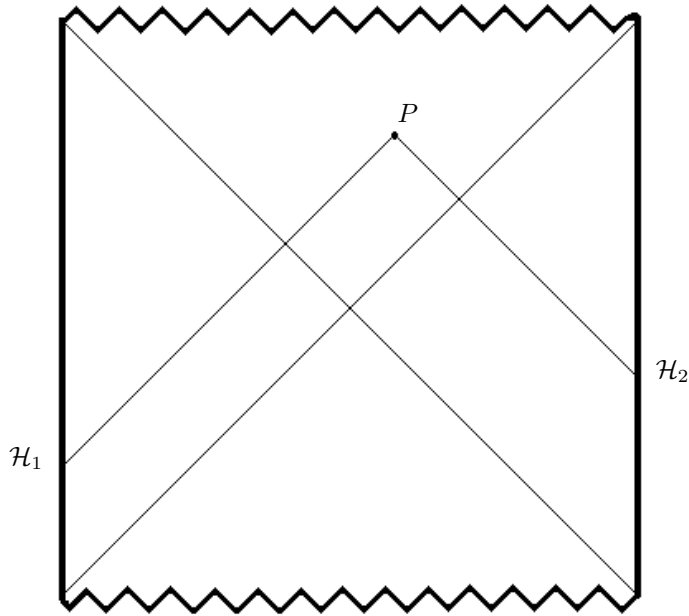}}{
\at(6\pscm;8\pscm){$P$} 
\at(10\pscm;3.9\pscm){${\cal H}_2$}
\at(-0.5\pscm;2.5\pscm){${\cal H}_1$} }}
\caption{The Penrose diagram for the maximally extended
Schwarzchild-AdS solution. The thicker vertical lines represent the
holographic screens. Holography tells us that a state $P$ inside the
black hole can be immersed into either Hilbert space ${\cal H}_1$ or
${\cal H}_2$. }
\end{center}
\end{figure}

On the other hand, it is usual for screens to form during gravitational
collapse. Usually,  however, for  normal and anti-trapped  regions the
projection into past screens is  not altered, and then the details of
the collapse can still be encoded into the screen at the
boundary. This can be illustrated by following the past null
projection of the point $P$ in Fig. 1. For instance, the  problem  of
gravitational collapse in  AdS was  studied in several  instances in
the  context of AdS/CFT,  and corresponds in the  dual theory  to  the
evolution  of excited states  into one or more states in a thermal
ensemble \cite{collapse}. The entropy of the black hole shows as the
degeneracy of the thermal ensemble corresponding to it.

\begin{figure}[ht]
\begin{center}
\mbox{\psannotate{\psboxto(10cm;0cm){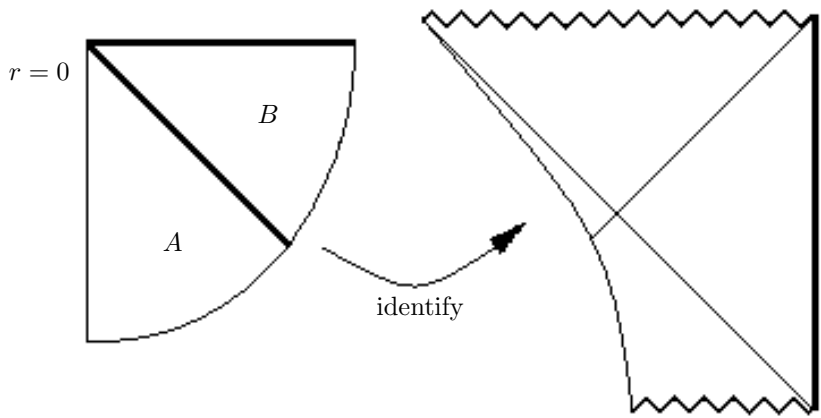}}{
\at(3.9\pscm;1.25\pscm){identify}
\at(-0.5\pscm;4\pscm){$r=0$}
\at(1.2\pscm;2\pscm){$A$}
\at(2.2\pscm;3.5\pscm){$B$}
}}
\caption{The birth of a child-universe in AdS space. A
domain wall separates the expanding region (to the left) to the vacuum
Schwarzchild solution (right). The holographic screens are represented
in thicker lines.}
\end{center}
\end{figure}

But ultimately the details of the collapsing region depend heavily on
the model studied. One can cook up models in which a cosmological type
of horizon is formed inside the region under collapse, which, in its
extreme view,  can create causally disconnected patches of space-time
\cite{guth2}, Fig. 2. Our goal is a little more modest: while the uniqueness
of the compactifications of 11-dimensional supergravity prevents the
creation of ``child-universes,'' we would still hope that anti-trapped
surfaces could be created in a meta-stable state of the theory. Thus,
while untimately the scenario from Fig. 2 would decay into the
Schwarzchild-AdS black hole, it is also clear that not all
states inside the bubble would allow for a unique description at 
spatial infinity. For instance, region $B$ would allow for a spatial
boundary description by ultimately decaying into AdS space, but the
degrees of freedom in region $A$ are projected onto the null
screen. The past projection cannot be used since in this scenario
the past directed null geodesics are not complete \cite{guthfarhi}.

The puzzle relies on this simple fact: by considering a bubble of these
states into the usual AdS vacuum of supergravity compactifications,
one might create an anti-trapped surface inside a trapped one. As the
system evolves, the meta-stable state would decay and form a
singularity. So future projection of the degrees of freedom inside the
black hole are forbidden. But now degrees of freedom inside the bubble
and behind the horizon cannot be projected out of the black-hole
either, like the region $B$ in Fig. 2. If this construct can be
implemented, the description of the
black-hole in the boundary theory would be by means of a {\em density
matrix}, and not by a wave-function. The density matrix would encode
the fact that the degrees of freedom representing the meta-stable
state are not all (fore)seen by an observer at spatial infinity. 

This interpretation can perhaps be seen more clearly from the Kruskal
extension of the Schwarzchild-AdS solution (Fig. 1.) A
state localized in a region corresponding to a point $P$ inside the
black hole region can be encoded in either screen, ${\cal H}_1$ or
${\cal H}_2$. That means that in spite of the fact that the dual
theory, say in ${\cal H}_1$, is still able to determine the state in
$P$ uniquely using its Hilbert space, the evolution of that state will
depend also on information coming from the other leaf of the universe,
which is encoded by Hilbert spaces in the other screen, such as ${\cal
H}_2$. In other words, the total Hilbert space of space-time is not
nicely separated into the sum of two spaces, one for each screen. Any
dual description of the evolution of the state in $P$ would be imcomplete
since the theory is missing part of the Hilbert space. The dual theory
then sees the state as a density matrix.

\section{Deformations of the background}

Let us consider first the D1-D5 system. One starts from the low energy
effective Lagrangian of type IIB\footnote{For conventions and
background, see \cite{polchinski}} in Einstein metric:  \be
S_{\mbox{\scriptsize IIB}}=\frac{1}{g_s^2l_s^8}\int d^{10}x
\sqrt{-g}\left( R-\frac{1}{2}\nabla_a \Phi \nabla^a \Phi-\frac{1}{24}
e^{\Phi}H_{abc}H^{abc}\right)+\ldots \ee with an ellipsis for all
fields not turned on here.

The background described is the usual: there are $Q_1$ D1 branes and
$Q_5$ D5 branes, sources for the $H$ field. The solution for the
Einstein metric is: \be
\begin{array}{c}
ds_E^2=H_1^{-3/4}H_5^{-1/4}\eta_{\mu\nu}dx^\mu
dx^\nu+H_1^{3/4}H_5^{1/4}dx^idx^i+\frac{H_1^{1/4}}{H_5^{1/4}}dx^mdx^m;
\\ H_1=1+\frac{r_1^2}{r^2},~~~r_1^2=\frac{(2\pi )^2g_sQ_1l_s^6}{V_4};
\\ H_5=1+\frac{r_5^2}{r^2},~~~r_5^2=g_s Q_5 l_s^2. \\
\end{array}
\label{sol1}
\ee with $\mu$, $\nu$ parallel to all branes, $m$ tangent to the D5
branes but transverse to the D1, and $i$ transverse to all branes. To
keep the total energy finite, we will compactify the $m$ coordinates
on a four torus, with volume $V_4$.

Thus the six dimensional infinte space of the solution (\ref{sol1})
has two distinct regions: for $r\rightarrow\infty$ is assymptotically
flat, whereas in the limit $r\rightarrow 0$ it factorizes into
$\mbox{AdS}_3\times S^3$. The transition is at $r\approx r_1,~r_5$. By
taking the {\em near horizon} limit - $r\rightarrow 0$, one can single
out the $\mbox{AdS}_3\times \mbox{S}^3$ portion. In this limit, the
radius of curvature for both the $\mbox{AdS}_3$ and $\mbox{S}^3$ is of
order $Q_1Q_5$. The size of the torus, however is much smaller, being
of order $Q_1/Q_5$. This means that we have an effective
six-dimensional supergravity on $\mbox{AdS}_3\times \mbox{S}^3$ with
matter fields arising from the compactification. From the
10-dimensional point of view, we are considering low energy
excitations near the branes, which, due to the non-flat background,
never make it to the flat region. 

As far as the dynamics around this region is concerned, the dilaton
and the Kaluza-Klein modes will behave as matter fields in
$\mbox{AdS}_3$. We will restrict our attention to fluctuations in the
$S^3$ radius and the $T^4$ volume, disregarding higher and stringy
corrections. This system, however, contains some small compactified
dimensions, the torus, and some stringy states may be excited during
the evolution of this system. Taking them into account would provide 
a more through analysis of the system, but will not be done here.
Writing then the ansatz for the metric:

\be ^{10}g_{ab}=\tilde{g}^{(3)}_{ab}\oplus
\tilde{h}^{(3)}_{a_1b_1}\oplus
\tilde{\delta}^{(4)}_{a_2b_2}=e^{-6\alpha-8\beta} g_{ab}\oplus
e^{2\alpha}~h_{a_1b_1} \oplus e^{2\beta}~\delta_{a_2b_2} \ee with $h_{ab}$ and
$\delta_{ab}$ constant curvature metrics for $S^3$ and $T^4$ and
$\alpha$ and $\beta$ generic functions of the ``$\mbox{AdS}_3$''
coordinates $\{t,x,y\}$. The constraint of $Q_1$  electric and $Q_5$
magnetic charges determines $H$ completely as a function of $\alpha$
and $\beta$:

\be H = \frac{(2\pi l_s)^2Q_5}{l^3}
e^{-3\alpha}\varepsilon+\frac{(2\pi g_s l_s)^2 Q_1 l_s^4}{V_4}
e^{-3\alpha-4\beta}\varepsilon_{S^3} \ee with $\varepsilon_g$
the volume form in $\{t,x,y\}$ associated with $g_{ab}$, and
$\varepsilon_{S^3}$ the volume form of the 3-sphere with metric $h_{a_1b_1}$.
It will be useful to set the radius of the three-sphere to
$l^2=l_s^2\frac{g_sl_s^4}{V_4}\sqrt{Q_1Q_5}$ so $\alpha = \beta =0$
will correspond to the vacuum solution.  The reduced equations are
(see Appendix):

\be
\begin{array}{c}
R_{ab}= 12
\nabla_a\alpha\nabla_b\alpha+24\nabla_{(a}\alpha\nabla_{b)}\beta + 20
\nabla_a\beta\nabla_b\beta + \frac{1}{2}\nabla_a\Phi\nabla_b\Phi -
~~~~~~~~~~~~~~~ \\ ~~~~~~~~~~~~~~~~~~~~~~~~~~~~~~~~~~~~~~~~~~ -
\frac{g_{ab}}{l^2}e^{-12\alpha-8\beta} \left( 6
e^{2\alpha}-2e^{-4\beta-\Phi}-2 e^\Phi\right) \\
\nabla_a\nabla^a\alpha = \frac{1}{l^2}\left(
4e^{2\alpha}-e^{-4\beta-\Phi}-3e^\Phi\right) e^{-6\alpha-4\beta} \\
\nabla_a\nabla^a\beta = \frac{1}{l^2}\left( e^\Phi -
e^{-4\beta-\Phi}\right) e^{-6\alpha-4\beta} \\ \nabla_a\nabla^a\Phi =
\frac{2}{l^2}\left( e^\Phi - e^{-4\beta-\Phi}\right)
e^{-6\alpha-4\beta} \\
\end{array}
\label{eqns1}
\ee From above one sees the flat direction $\gamma=2\beta-\Phi$,
having to do with the variation of the fields that leaves the
effective 6-dimensional Newton's constant invariant. Shifting $\gamma$
amounts then to shifting the minimum of the potential, which in turn
can be trivialized by changing $l$. From above it turns out that it's
also desirable to define $\varphi=\alpha+\beta$. The independent
equations of motion in (\ref{eqns1}) are derivable from an action:

\be S_{\mbox{eff}}=\frac{1}{^3G_N}\int
d^3x\sqrt{-g}\left(R-3\nabla_a\varphi
\nabla^a\varphi-4\nabla_a\beta\nabla^a\beta
+\frac{6}{l^2}e^{-4\varphi}-\frac{4}{l^2}e^{-6\varphi}\cosh
4\beta\right)  \ee plus a term $\nabla_a\gamma\nabla^a\gamma$ to
account for the dilaton equation. One notes in the equations of
motions (\ref{eqns1}) and the Lagrangian  above the symmetry
$\beta\rightarrow -\beta$, a manifestation of U-duality along the
branes directions. Note that the matter fields don't necessarily
satisfy the strong energy condition.

\section{Inflation?}

The exponential growth of scales was proposed about 20
years ago \cite{guth1} to explain the assymptotic flatness and the
horizon problem of the recent universe. Along with the idea, a simple
system was proposed where an unstable excited state of a scalar field
would drive the exponential or polynomial growth of spatial
scales. the idea behind is that, if the 
spatial scales grow too fast \cite{halliwell} (faster than the proper
time), two nearby observers won't be able to see each other, if the
expansion lasts for enough time. A cosmological horizon would then be formed.

As usual, one slices up space-time into homogeneous surfaces of
constant time, whose flow is represented by a vector field
$\xi^a=\nabla^a t$. Orbits of $\xi^a$  are geodesics
($\xi^a\nabla_a\xi^b=0$), and $\xi^a\xi_a$ is normalized to $-1$. This
allows us to interpret $\nabla_a\xi_b$ as the time derivative of the
spatial metric ${{\cal L}}_{\xi}h_{ab}=\dot{h}_{ab}$\footnote{See, for
instance, Appendix E in \cite{wald}}. Because space is assumed to be
homogeneous, the spatial metric $h_{ab}$ has constant curvature, and
the only allowed evolution is a dilation
$\nabla_a\xi_b=\dot{h}_{ab}=\theta h_{ab}$. The dynamical system
(\ref{eqns1}) will then inflate if there  is a critical point for
which the expansion factor $\theta$ is positive. The equation for
$\theta$ is then just the Raychaudhuri equation \cite{wald}: \be
\dot{\theta}=-\frac{1}{2}\theta^2-\xi^a R_{ab}
\xi^b=-\frac{1}{2}\theta^2-3\dot{\varphi}^2-4\dot{\beta}^2-\frac{2}{l^2}\left(
3 e^{-4\varphi}-2e^{-6\varphi}\cosh 4\beta\right)
\label{ray}
\ee Using these coordinates, the equations for $\phi$ and $\beta$ are

\be
\begin{array}{c}
\ddot{\varphi}=
-\theta\dot{\varphi}-\frac{4}{l^2}e^{-4\varphi}+\frac{4}{l^2}e^{-6\varphi}
\cosh 4\beta  \\ \ddot{\beta}=
-\theta\dot{\beta}-\frac{2}{l^2}e^{-6\varphi}\sinh 4\beta \\
\end{array}
\end{equation}

If the system is now placed at a point where $4\beta \gg 2\varphi \sim
0$, the $\beta$ term will dominate in the potential. After a
reparametrization $\frac{d\tau}{dt}=e^{-3\varphi+2\beta}$, the system
can be written as:

\be
\begin{array}{c}
\tilde{\theta}'\simeq -\frac{1}{2}\tilde{\theta}^2+3\varphi
'\tilde{\theta}-2\beta  '\tilde{\theta}-3\varphi '^2-4\beta
'^2+\frac{2}{l^2} \\ \varphi '' \simeq 3\varphi '^2-2\beta '\varphi
'-\tilde{\theta}\varphi ' +\frac{2}{l^2} \\ \beta '' \simeq 3\beta
'\varphi '-2\beta '^2-\tilde{\theta}\beta '-\frac{1}{l^2}
\end{array}
\ee where $\tilde{\theta}=\theta e^{3\varphi-2\beta}$ and the primes
denote differentiation with respect to $\tau$.

As it turns out the system above does have a critical point: \be
\tilde{\theta}=\frac{4\sqrt{2}}{l},\mbox{    }\beta
'=-\frac{1}{2\sqrt{2}l}, \mbox{    }\varphi '=\frac{1}{\sqrt{2}l} \ee

With the initial conditions, $\beta(0)=\beta_0$ and $\phi(0) = 0$, the
approximate solution  for $t\ll l e^{2\beta_0}$ is:
\be
\begin{array}{c}
\theta = \frac{\frac{4\sqrt{2}}{l}
e^{2\beta_0}}{1+\frac{2\sqrt{2}}{l}e^{2\beta_0}t}\\ \beta =
\beta_0-\frac{1}{8}\ln \left(1+2\sqrt{2}e^{2\beta_0}\frac{t}{l}
\right) \\ \varphi =
\frac{1}{4}\ln\left(1+2\sqrt{2}e^{2\beta_0}\frac{t}{l} \right) .
\end{array}
\ee 
As $t$ approaches $l e^{2\beta_0}$ the number of e-foldings
$\frac{1}{2}\int \theta dt$ grows as $4\beta_0$.
Observers following the flow of time $\xi^a$ will then see an
expansionary phase for an arbitrarily long time. After some number of
e-foldings, however, the system will relax to its unique ground state:
$\mbox{AdS}_3$.

Information about the space slices can be gathered by decomposing the
curvature: $^3R= {^2R}+\frac{3}{2}\theta^2+2\dot{\theta}$ and taking
the trace in (\ref{eqns1}). Using the equation of motion for
$\dot{\theta}$ (\ref{ray}) one arrives at the Hamiltonian constraint:
\be
{^2R}=-\frac{1}{2}\theta^2+3\dot{\varphi}^2+4\dot{\beta}^2-\frac{1}{l^2}
(6e^{-4\varphi}-4e^{-6\varphi}\cosh 4\beta ). 
\ee 
At the critical point, and at early stages $t\ll l e^{2\beta_0}$, we have: 
\be
^2R\simeq -\frac{12}{l^2}e^{-6\varphi+4\beta}=
-\frac{12}{l^2}\frac{e^{4\beta_0}}{\left(1+\frac{2\sqrt{2}}{l}
e^{2\beta_0}t\right)^2},  
\ee 
so, although the system initially has a
large spatial curvature, it actually ``opens up'' $^2R\rightarrow 0$
as a result of the expansion. The spatial curvature is, however,
negative definite, and a cosmological horizon would still not be
formed.  In fact, the geometry for intermediate times is not unlike
the radiation-filled hyperbolic FRW universe. Despite having
non-trivial exponential potentials for the scale factors of
$\mbox{S}^3$ and $\mbox{T}^4$, the conditions for inflation are not met.

The resulting 10-dimensional geometry is computed by reversing the
definition of $\varphi$ and the conformal transformations relating the
six-dimensional and the ten-dimensional metrics:

\be
ds^2_{(10)}=-d\nu^2+K\nu^{2}t_{ij}dx^idx^j+\ldots
\ee
with $t_{ij}$, ($i,j=1,2$) a metric of constant negative curvature, K a
constant depending on the initial conditions and the metric of the
sphere and the torus were omitted. Note
that the ten dimensional causal structure is the same as the
three-dimensional one, due to the symmetry of the internal space 
and the fact that the two metrics are related by a conformal
transformation. 

\section{Other Examples}

\subsection{$\mbox{AdS}_5\times\mbox{S}^5$}

Much the same way the $\mbox{D1-D5}$ system gives rise to the
$\mbox{AdS}_3\times\mbox{S}^3\times\mbox{T}^4$ space, the
near-horizon geometry of a number of $D3$ branes is
$\mbox{AdS}_5\times\mbox{S}^5$. We turn now to study deformations around
this background. The situation here is {\it a priori} a bit different
from the previous example in which there are no small dimensions: the
radius of the five-sphere is the same as the radius of
$\mbox{AdS}_5$. There is then some justification in considering
perturbations of the radii of each direct summand of space-time. The
system is expected to behave more ``gravitationally'' than ``stringy''
at the moderate energies we are dealing with.

One starts from the Ansatz: 
\be 
^{10}g_{ab}=\tilde{g}^{(5)}_{xy}\oplus
\tilde{h}^{(5)}_{a_1b_1} = e^{-\frac{10}{3}\alpha}~{g}_{xy} \oplus
e^{2\alpha}~{h}_{a_1b_1} 
\ee 
The 3-branes act as electric and magnetic sources for the $F_5$ field: 
\be
\int_{S^5}{^{10\star}}F_5=N=\int_{S^5}F_5 
\ee  
By the condition of duality (see, for instance, \cite{polchinski}). Then 
\be
F_5=N(e^{-5\alpha}\tilde{\varepsilon}_{S^5}+ie^{-5\alpha}
\tilde{\varepsilon}_{\tilde{g}}) 
\ee with $\tilde{\varepsilon}$ being the
volume form associated with the full metric (before the scaling
factors have been extracted). The stress-energy is then: 
\be
T_{xy}=-\frac{2}{3l^2} e^{-10\alpha}\tilde{g}_{xy} 
\ee 
\be
T_{a_1b_1}=+\frac{2}{3l^2} e^{-10\alpha}\tilde{h}_{a_1b_1},
\ee 
with
$l$ defined so that $\alpha =0$ correspond to the vacuum of the
theory. Then $l$ has some factors of the coupling and the number of
3-branes. The equations of motion are: 
\be
R_{xy}=\frac{40}{3}\nabla_x\alpha\nabla_y\alpha -
\frac{5}{3}\nabla^2\alpha~g_{xy}-\frac{2}{3l^2}
e^{-\frac{40}{3}\alpha}~g_{xy} 
\ee 
and 
\be
\nabla^2\alpha=\frac{2}{3l^2}e^{-\frac{22}{3}\alpha} -\frac{2}{3l^2}
e^{-\frac{40}{3}\alpha}.
\ee  
Defining the unit, future directed,
time vector $\xi^a$, one can write the Raychaudhuri equation for the
system: 
\be \dot{\theta}=-\frac{1}{4}\theta^2-\xi^aR_{ab}\xi^b=
-\frac{1}{4}\theta^2- \frac{40}{3}\dot{\alpha}^2- \frac{2}{3l^2}\left(
\frac{5}{3}e^{-\frac{22}{3}\alpha} -
\frac{2}{3}e^{-\frac{40}{3}\alpha}\right),  
\ee 
where
$\theta=\nabla_a\xi^a$ is the expansion factor, related to the Hubble
factor by $H=\frac{\dot{a}}{a}=\frac{1}{4}\theta$. Dots mean
differentiation with respect to proper time, which happens to be the
affine parameter of integral curves of $\xi^a$. The equation for
$\alpha$ is  
\be
-\ddot{\alpha}-\theta\dot{\alpha}=\frac{2}{3l^2}e^{-\frac{22}{3}\alpha}
-\frac{2}{3l^2} e^{-\frac{40}{3}\alpha},
\ee 
on the supposition that
spatial derivatives are small, and hence space is homogeneous.

The stable critical point of the system above, parametrized with
$\frac{d\tau}{dt}= e^{-\frac{20}{3}\alpha}$, for $\alpha \ll 0$ is: \be
\begin{array}{c}
\tilde{\theta}=\theta
e^{-\frac{20}{3}\alpha}=\frac{8}{3}\sqrt{\frac{10}{3}} \\ \alpha
'=\frac{d\alpha}{d\tau}=\frac{1}{\sqrt{30}}
\end{array}
\ee Then $\frac{a(t)}{a_0}=\exp\left( \frac{1}{4}\int
\theta\right)\sim t$, the evolution is FRW-like and no cosmological
horizon is formed.

\subsection{$\mbox{AdS}_7\times\mbox{S}^4$}

This space arises as the near horizon limit of the eleven dimensional
supergravity (11-SUGRA) solution for a system of M5
branes\footnote{See, for instance, Chapter 6 in \cite{adscft} for a
review.}. The AdS/CFT conjecture here is quite interesting since
neither the string theory on this backgroung (the ``little string
theory'') neither its dual conformal field theory (the $A_{N-1}$
$(2,0)$ 6-dimensional SCFT) are known enough to fully test the results
predicted. As stressed in the second section, knowing whether horizons
form in excitations of the theory is then important if one is to
encode the information inside $\mbox{AdS}_7$ in its boundary, by
Bousso's prescription.

The M5 branes are magnetic sources for the 11-SUGRA's antisymmetric
4-tensor field:

\be
\int_{S^4}F_4=N .
\ee
Note that, with this solution \cite{freund}, the Chern-Simons term
vanishes and doesn't contribute to the equations of motion.

So the energy-momentum tensor, written in terms of the eleven
dimensional metric $\tilde{g}^{(11)}_{ab}=\tilde{g}^{(7)}_{xy}\oplus
~\tilde{h}^{(4)}_{a_1b_1}= e^{2\psi}~g^{(7)}_{xy}\oplus
e^{2\alpha}~h^{(4)}_{a_1b_1}$ is:
\be
\begin{array}{c}
\tilde{T}_{xy}=-\frac{1}{4}N^2e^{-8\alpha}\tilde{g}_{xy} \\
\tilde{T}_{a_1b_1}=\frac{1}{4}N^2e^{-8\alpha}\tilde{h}_{a_1b_1}.\\
\end{array}
\ee

From the Appendix, $\psi=-\frac{4}{5}\alpha$, and then one can readily
write the Einstein equations:

\be
R_{xy}=\frac{36}{5}\nabla_x\alpha\nabla_y\alpha -
\frac{4}{5}\nabla^2\alpha~g_{xy}-\frac{1}{l^2}
e^{-\frac{48}{5}\alpha}~g_{xy} 
\ee 
and
\be
\nabla^2\alpha=\frac{2}{l^2}e^{-\frac{18}{5}\alpha} -\frac{2}{l^2}
e^{-\frac{48}{5}\alpha} 
\ee

The dynamical system for $\alpha\ll 0$ is approximated by:
\be
\begin{array}{c}
\tilde{\theta}'-\frac{24}{5}\alpha'\tilde{\theta}+
\frac{1}{6}\tilde{\theta}^2+\frac{36}{5}\alpha'^2 = \frac{3}{5l^2} \\
\alpha''-\frac{24}{5}\alpha'^2+\tilde{\theta}\alpha'=\frac{2}{l^2} \\
\end{array}
\ee
with $\tilde{\theta}=\theta e^{\frac{24}{5}\alpha}$ and
$\alpha'=\frac{d\alpha}{d\tau}=\frac{d\alpha}{dt}e^{-\frac{24}{5}\alpha}$.
The system has a critical point at:
\be
\tilde{\theta}=\frac{24\sqrt{3}}{5l},\;\alpha'=\frac{1}{2\sqrt{3}l}
\ee
with $a(t)\approx t$.

\section{Discussion}

The perspective of disjoint holographic screens seems both interesting
and mysterious. For asymptotically AdS spaces it would mean
that the description we have of a gravitational state, like a black
hole, in terms of the boundary theory is at best in terms of a density
matrix. Bousso \cite{bousso} circumvented the problem by
distinguishing between the spatial-infinity conformal boundary (dual)
theory and the holographic one. The dual theory would only know about
those patches of space-time whose degrees of freedom could be
projected to spatial infinity, whereas the holographic theory would
assign Hilbert spaces to each point of each screen and give
correlation functions between them with no further distinction. One
would then like to embed this construction in string theory, and
possibly give it a more precise account.

However, it seems that solutions with disjoint screens  cannot be
found as distortions of the simple compactifications of
string/M-theory in spaces with some symmetry. The reason lies on the
construction of those reduced systems, which, despite coming from
string sources, are weakly coupled enough to allow for a supergravity
description of the matter fields.  The constraints of 11 dimensional
supergravity still hold in the reduced system. One can see that from
the fact that the 11-dimensional metric is related to the reduced
metric by a conformal transformation. A horizon formed by inflation in
the latter would also mean a horizon in the former, which violates the
(11 dimensional) strong energy condition. If string theory does have
such solutions and holography is expected to hold, perhaps by taking
the near-horizon limit one is constraining the set of initial
conditions allowed by the system. The dual boundary theory would not
be able to describe the set of conditions which would form a screen in
the bulk.

So it seems that a further knowledge of string effects at moderate
energies is necessary. A good laboratory for this might again be the
$\mbox{D1-D5}$ system, with its small compactified dimensions. 
Combinations of massive fields and exponential potentials could
prove to be just the ingredients necessary to spawn inflation
\cite{halliwell}. Another way to proceed is to look at backgrounds
with less supersymmetry. Intersecting branes are an example, where the
matter fields do not arise from the gravity multiplet. On the other
hand, by having less supersymmetry one allows for non-perturbative
effects \cite{bbs}, which may generate potentials for moduli which do
not have a direct geometrical interpretation in terms of ten or eleven
dimensional supergravity. More work is indeed necessary.

As this paper was in its final phase of preparation I became aware of
another work by Hellerman {\it et al.} \cite{susskindetalii} which
arrived at similar conclusions.
 
\section*{Acknowledgements}

I would like to thank  Emil Martinec for the idea, guidance 
and especially patience. I would also like to thank R. M. Wald and
P. G. O. Freund for rewarding discussions.

\section*{Appendix: Kaluza-Klein reduction and scale deformations}

The most general metric used here has the form:
\be 
\tilde{g}^D_{ab}=e^{2\psi}~{g^{(p)}_{xy}}\oplus
e^{2\alpha}~{h^{(m)}_{a_1b_1}}\oplus e^{2\beta}~{h^{(n)}_{a_2b_2}}
\ee
of a $10$ or $11$ dimensional space compactified on $\mbox{S}^m$ and
$\mbox{T}^n$. We will use the axioms of the covariant derivative to
relate the Ricci tensor $\tilde{R}_{ab}$ associated with
$\tilde{g}^D_{ab}$ with the corresponding $R_{ab}$ obtained as
$\psi=\alpha=\beta=0$. Beginning with the Christoffel symbols:

\be {C^x}_{yz}=\nabla_y\psi~\delta^x_z+
\nabla_z\psi~\delta^x_y-\nabla^x\psi~g_{yz}, \ee
\be {C^x}_{a_1b_1}=-\frac{1}{2}e^{-2\psi}\nabla^x(e^{2\alpha})=
-e^{-2\psi+2\alpha}\nabla^x\alpha ~h_{a_1b_1}, \ee \be
{C^{a_1}}_{xb_1}=\frac{1}{2}e^{-2\alpha}\nabla_x(e^{2\alpha})
\delta^{a_1}_{b_1}=\nabla_x\alpha~\delta^{a_1}_{b_1}, \ee with
analogous expressions for $\beta$. All indices will from here on be
raised with $g^{(p)}_{xy}$. Proceeding by computing the parcels in the
definition of the curvature tensor:
\begin{eqnarray*}
& {C^a}_{ax}=p\nabla_x\psi+m\nabla_x\alpha+n\nabla_x\beta , & \\ &
\nabla_b{C^b}_{xy}=\nabla_z{C^z}_{xy}=2\nabla_x\nabla_y\psi-g_{xy}
\nabla^2\psi , & \\
\end{eqnarray*}
$$
\begin{array}{c} 
{C^a}_{xy}{C^d}_{da}= 2p\nabla_x\psi\nabla_y\psi
-pg_{xy}(\nabla\psi)^2 + 2m\nabla_{(x}\psi \nabla_{y)}\alpha
-mg_{xy}\nabla_z \psi\nabla^z\alpha +\\ \;\;\;\; +
2n\nabla_{(x}\psi \nabla_{y)}\beta -ng_{xy}\nabla_z
\psi\nabla^z\beta ,  \\
\end{array}
$$
\begin{eqnarray*}
{C^a}_{xb}{C^b}_{ya} & = &
{C^z}_{xt}{C^t}_{yz}+{C^{a_1}}_{xz}{C^z}_{ya_1}+{C^{a_2}}_{xz}{C^z}_{ya_2}
\\  & = & (p+2)\nabla_x\psi\nabla_y\psi-2g_{xy}~(\nabla\psi )^2
+m\nabla_x\alpha\nabla_y\alpha+n\nabla_x\beta\nabla_y\beta .\\
\end{eqnarray*}

Then we obtain

\be
\begin{array}{c}
\tilde{R}_{xy}=R_{xy}-(p-2)\nabla_x\nabla_y\psi-m\nabla_x\nabla_y\alpha
-n\nabla_x\nabla_y\beta -g_{xy}\nabla^2\psi+ \\
\;\;\;\;
(p-2)\nabla_x\psi\nabla_y\psi-(p-2)g_{xy}~(\nabla\psi )^2+
2m\nabla_{(x}\psi\nabla_{y)}\alpha - 
mg_{xy}~\nabla_z\psi\nabla^z\alpha + \\ \;\;\;\;
 2n\nabla_{(x}\psi\nabla_{y)}\beta -
ng_{xy}~\nabla_z\psi\nabla^z\beta -m\nabla_x\alpha\nabla_y\alpha
-n\nabla_x\beta \nabla_y\beta ,\\
\end{array}
\ee 
and

\be
\begin{array}{c}
\tilde{R}_{a_1b_1}=R_{a_1b_1}-e^{-2\psi+2\alpha}\left[\nabla^2\alpha
+ \left((p-2)\nabla^x\psi+m\nabla^x\alpha+n\nabla^x\beta\right)
\nabla_x\alpha\right] h_{a_1b_1} ,\\
\tilde{R}_{a_2b_2}=R_{a_2b_2}-e^{-2\psi+2\beta}\left[\nabla^2\beta +
\left((p-2)\nabla^x\psi+m\nabla^x\alpha+n\nabla^x\beta\right)
\nabla_x\beta \right] h_{a_2b_2}.
\end{array}
\ee

One sees from above that choosing
$\psi=-\frac{m}{p-2}\alpha-\frac{n}{p-2}\beta$ one can not only make
the two-derivative terms in the reduced Ricci tensor vanish but also
the non-linear terms in the internal curvature. This choice is called
the Einstein frame since then the  energy-momentum tensor for the
extra ``matter'' fields $\alpha$, $\beta$, which can be constructed
from the equation for $\tilde{R}_{xy}$ above via the Einstein equation,
will be conserved.

\end{document}